\documentclass[reprint,
superscriptaddress,
 amsmath,amssymb,
prb,
floatfix,
]{revtex4-1}

\usepackage{graphicx}
\usepackage{dcolumn}
\usepackage{bm}

\begin{document}

\preprint{APS/123-QED}

\title{N-nH complexes in GaAs studied at the atomic scale by cross-sectional scanning tunneling microscopy}

\author{D. Tjeertes}
\email{d.tjeertes@tue.nl}
\author{T.J.F. Verstijnen}%
\affiliation{%
 Department of Applied Physics, Eindhoven University of Technology, P.O. Box 513, 5600 MB Eindhoven, The Netherlands
}%

\author{A. Gonzalo}
\author{J.M. Ulloa}
\affiliation{
 Institute for Systems based on Optoelectronics and Microtechnology (ISOM), Universidad Politécnica de Madrid, Ciudad Universitaria s/n, 28040 Madrid, Spain
}%

\author{M.S. Sharma}
\author{M. Felici}
\author{A. Polimeni}
\affiliation{%
Dipartimento di Fisica, Sapienza Università di Roma, I-00185 Rome, Italy
}%

\author{F. Biccari}
\author{M. Gurioli}
\affiliation{Department of Physics and Astronomy,and LENS, University of Florence, I-50019 Sesto Fiorentino (FI), Italy}

\author{G. Pettinari}
\affiliation{National Research Council, Institute for Photonics and Nanotechnologies (IFN-CNR), I-00156 Rome,
Italy}

\author{C. \c{S}ahin}
\affiliation{Department of Physics and Astronomy, Optical Science and Technology Center, University of Iowa, Iowa City, Iowa 52242, USA}

\author{M.E. Flatt\'{e}}
\affiliation{Department of Physics and Astronomy, Optical Science and Technology Center, University of Iowa, Iowa City, Iowa 52242, USA}
\affiliation{%
 Department of Applied Physics, Eindhoven University of Technology, P.O. Box 513, 5600 MB Eindhoven, The Netherlands
}%

\author{P.M. Koenraad}%
\affiliation{%
 Department of Applied Physics, Eindhoven University of Technology, P.O. Box 513, 5600 MB Eindhoven, The Netherlands
}%

\date{\today}

\begin{abstract}
Hydrogenation of nitrogen (N) doped GaAs allows for reversible tuning of the bandgap and the creation of site controlled quantum dots through the manipulation of N-nH complexes, N-nH complexes, wherein a nitrogen atom is surrounded by n hydrogen (H) atoms. Here we employ cross-sectional scanning tunneling microscopy (X-STM) to study these complexes in the GaAs (110) surface at the atomic scale. In addition to that we performed density functional theory (DFT) calculations to determine the atomic properties of the N-nH complexes. We argue that at or near the (110) GaAs surface two H atoms from N-nH complexes dissociate as an H$_2$ molecule.
We observe multiple features related to the hydrogenation process, of which a subset is classified as N-1H complexes. These N-1H related features show an apparent reduction of the local density of states (LDOS), characteristic to N atoms in the GaAs (110) surface with an additional apparent localized enhancement of the LDOS located in one of three crystal directions. N-nH features can be manipulated with the STM tip. Showing in one case a switching behavior between two mirror-symmetric states and in another case a removal of the localized enhancement of the LDOS. The disappearance of the bright contrast is most likely a signature of the removal of an H atom from the N-nH complex. 
\end{abstract}

\maketitle

\section{Introduction}
    Introducing a small ($<$5\%) amount of nitrogen (N) in III-V semiconductor materials results in a drastic reduction of their bandgap. For N-doped GaAs  this results in a bandgap reduction of up to 600 meV, as shown by calculations and experiments \cite{Weyers1992RedLayers,Shan1999BandAlloys,Shan1999EffectAlloys}. This strong reduction of the bandgap allows for N-doped GaAs compounds that reach the 1.3 and 1.55 $\mu$m telecommunication windows. More recently the effect that hydrogen (H) ion irradiation has on these materials was discovered. When N-doped GaAs material is exposed to a flux of low-energy (100 eV) H ions, the bandgap reduction caused by the introduction of N is fully reversed \cite{Polimeni2001Effectmn1/,Baldassarri2001Hydrogen-inducedWells}. This restoration of the GaAs bandgap is caused by the formation of N-nH (n$\geq$2) complexes, wherein a nitrogen atom is surrounded by n hydrogen atoms. The hydrogen ions diffuse through the material in a very sharp front (5 nm/decade concentration) allowing fine control over the hydrogenation profile \cite{Trotta2009HydrogenNx}. This control has been exploited to create site-controlled quantum dots (QDs) with the use of a hydrogen opaque mask \cite{birindelli2014single,Felici2020BroadbandDots}. Moreover, site-controlled QDs could be created by local H removal, achieved by illuminating the sample with a scanning near-field optical microscope (SNOM) \cite{Biccari2018Site-ControlledIllumination}.
    
    N-nH complexes in GaAs can exist in multiple configurations. N-1H complexes were shown to be stable if the H atom resides on a bond, centered between an N and Ga atom \cite{Bonapasta2002Structuremiy/mi}. The N-2H complexes are generally identified as the origin for the restoration of the bandgap of intrinsic GaAs, which has been shown through PL measurements \cite{Berti2007FormationNitrides}. The N-2H complexes also cause a relaxation of the tensile strain introduced by the N atoms, as shown through XRD measurements. On the contrary, N-3H complexes cause an inversion of the strain from tensile to compressive \cite{Bisognin2006Hydrogen-nitrogenStrain}. 
    The shape and orientation of the N-nH complexes has been studied through infrared absorption, nuclear reaction analysis and simulations \cite{Bonapasta2002Structuremiy/mi,Jiang2004b,Ciatto2005Nitrogen-hydrogen/mrow,Ciatto2009LocalNitrides,Wen2010DetailedStress,Wen2010DetailedStress,Bisognin2008High-resolutionNitrides,Amidani2014ConnectionsAlloys}. The basic consensus of those studies is that the hydrogen irradiation breaks two of the four N-Ga bonds and terminates them with H on the N side, forming an N-2H complex. For N-3H complexes the third H atom resides in the vicinity of the N atom, but the exact location is not yet known. The N-nH complexes are stable at room temperature, but when annealing samples containing these complexes at temperatures above 250$^\circ$C, the N-3H complexes start to dissociate, leaving N-2H complexes. At 300$^\circ$C and above, the N-2H complexes begin to dissociate, leaving non-hydrogenated N atoms behind \cite{Berti2007FormationNitrides}. 
    
    N atoms in N-doped GaAs have already been successfully imaged with the use of cross-sectional scanning tunneling microscopy (X-STM) \cite{Ulloa2008StructuralMicroscopy,Plantenga2017SpatiallyGaAs,McKay2001ArrangementMicroscopy,McKay2001DistributionMicroscopy}. Here we present the X-STM work on hydrogenated N-doped GaAs where we observe features related to the hydrogenation of N-doped GaAs. 
    
    We describe a selection of the features in the (110) surface plane of hydrogenated N-doped GaAs in detail. In addition, density functional theory calculations where performed on N-nH complexes in the same surface to study their atomic properties. Based on these results we argue that at or near the (110) GaAs surface two H atoms from N-nH complexes dissociate as an H$_2$ molecule. The N-1H related X-STM features show an apparent reduction of the local density of states (LDOS), characteristic to N atoms in the GaAs (110) surface with an additional apparent localized enhancement of the LDOS located in one of three crystal directions. Furthermore, we describe the manipulation of some of these features with an STM tip. This results in one case in a switching behavior between two mirror-symmetric states and in another case in the removal of the localized enhancement of the LDOS. The removal of the localized enhancement of the LDOS is most likely a signature of the removal of an H atom from the N-nH complex. 

\section{Methods}
    Two N-doped GaAs structures were grown using molecular beam epitaxy (MBE) on n$^+$-doped (100) GaAs substrates. Nitrogen was provided by a radio-frequency plasma source. The first sample contains a 200 nm bulk layer of N-doped GaAs grown on top of a 250 nm intrinsic GaAs buffer layer. The N-doped layer contains 0.62\% N as estimated by x-ray diffraction (XRD). The second sample contains a 100 nm N-doped GaAs layer with 1\% N.
    
    Optimal hydrogen exposure conditions were simulated based on the models described by Trotta \textit{\textit{et al}} \cite{Trotta2009HydrogenNx}, before the exposure was performed. Room temperature photoluminescence (PL) measurements were performed before and after the exposure to ensure that the hydrogenation had been successful. Hydrogen exposure was performed using a Kaufman plasma source. 
    The first sample was passivated with an impinging H dose of $2.5\times10^{18}$ ions/cm$^2$ at 300$^\circ$C. This should provide a full hydrogenation of the sample with a mixture of N-2H and N-3H complexes. From now on we will refer to this sample as the ``non-annealed sample''.
    The second sample was fully hydrogenated at 300$^\circ$C with an impinging hydrogen plasma dose of $9\times10^{17}$ ions/cm$^2$ and afterwards annealed under vacuum conditions for 10 h at 250$^\circ$C to dissociate the majority of the N-3H complexes to N-2H. From now on we will refer to this sample as the ``annealed sample''. 

    For X-STM measurements, the samples are brought into the ultra-high vacuum STM chamber (typical pressure below $5\times10^{-11}$ mbar) and cleaved \emph{in situ}, revealing a (110) plane of the sample. This allows us to image a cross-section of the sample showing all the grown layers. All measurements were performed at 77 K in an Omicron LT-STM in constant current mode. STM tips were electro-chemically etched from poly-crystalline tungsten (W) wire and further tip preparation was done by sputtering with argon (Ar). 
    
    During STM measurements electrons can either tunnel from the sample into the tip, or from the tip into the sample, depending on the applied bias voltage. The condition where electrons tunnel from the sample into the tip is called filled-state imaging, since the electrons contributing to the tunnel current originate from the filled energy states of the sample. Conditions under which electrons tunnel from the tip into the sample are called empty-state imaging, since the electrons move from the tip into the empty energy states of the sample. For semiconductors at low temperatures with the Fermi level in the band gap, filled state imaging implies that the electrons originate from the valence band, while in empty state imaging they are injected into the conduction band. 
    N doping introduces a state in the conduction band \cite{Shan1999BandAlloys,Hjalmarson1980TheorySemiconductors}. By using empty-state imaging these states can be observed as an increase in the amount of available states to tunnel into during empty-state imaging at specific voltages.

    Atomic relaxation calculations shown in this study are performed using density functional theory (DFT) software, specifically Quantum Espresso version 6.4.1 \cite{Giannozzi2009QUANTUMMaterials,EnkovaaraAdvancedESPRESSO}. The four types of atoms considered in these calculations are Ga, As, N and H. The core electrons are treated within a pseudopotential method with ultrasoft pseudopotentials \cite{verstijnen2020pseudo} and exchange-correlation energy functional is computed with the local density approximation (LDA). To simulate surface behavior a 56 atom supercell of a (110) GaAs surface with an N impurity on an As site in the surface layer is considered. This supercell is 7 atomic layers thick and there is 1.2 nm of vacuum above the surface. In this surface supercell up to 3 H atoms are introduced near the N impurity to simulate the N-nH complexes. The cutoff energy of the plane-wave basis set is determined to be 100 Ry via convergence tests and a 2$\times$2$\times$1 Monkhorst-Pack k-point grid is used for the Brillouin zone integration. On these supercells relaxation calculations are then performed using a force convergence threshold of $1\times10^{-4}$ hartrees/bohr.

\section{Results and Discussion}
    Earlier X-STM studies on N-doped GaAs provide a complete description of the properties of isolated undecorated N atoms in or near the (110) surface of GaAs \cite{Ulloa2008StructuralMicroscopy,Plantenga2017SpatiallyGaAs,McKay2001ArrangementMicroscopy,McKay2001DistributionMicroscopy}. Under filled state (negative voltage) imaging conditions N atoms give rise to a depression of the (110) surface, with the size and depth of the depression indicating how far below the surface plane the N atom is located. In empty state imaging (positive voltage) N atoms show an anisotropic bright structure which is highly dependent on the tunneling voltage and the distance the atom resides from the (110) cleaved surface \cite{Plantenga2017SpatiallyGaAs}\cite{Ishida2015DirectMicroscopy}. This information provides us with a way to determine the exact position of the N atoms with respect to the surface in the measurements. 
    
    \begin{figure}
            \centering
            \includegraphics{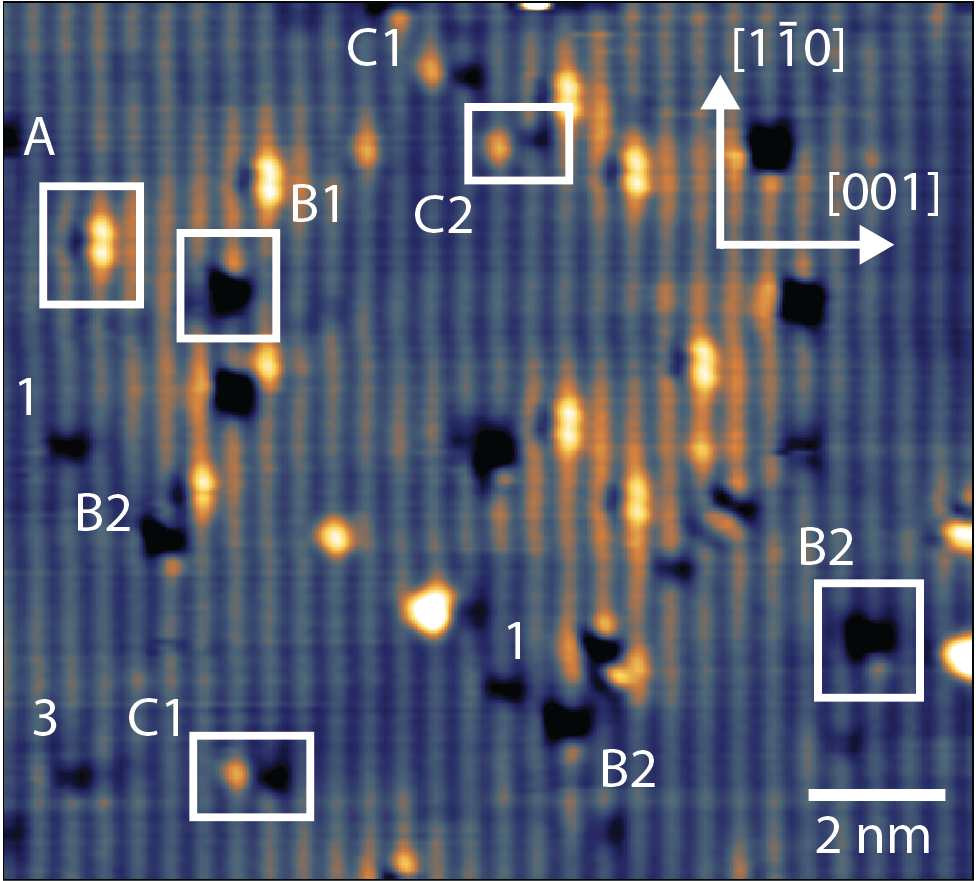}
            \caption{Filled state image of the non-annealed sample containing a variety of features obtained at a sample bias of -3.35 V and a tunnel current of 30 pA. The numbers 1 and 3 mark N atoms in the 1st (surface) and 3rd layer. The letters A mark Ga atoms, and B1/B2 and C1/C2 mark N-nH complexes. One instance of each of the features is bordered with a white rectangle for clarity.}
            \label{fig:features}
    \end{figure}
    
    A typical filled state measurement on the hydrogen irradiated N-doped GaAs layer of the non-annealed sample is shown in Fig. \ref{fig:features}. Running from top to bottom in bright lines are the imaged As atoms. In addition to that, we can observe dark and bright features. Marked with 1 and 3 are non-hydrogenated N atoms, where the numbers indicate their position with respect to the surface (with 1 being the surface layer and 3 two monolayers below the surface). Features marked with A, B1/B2 and C1/C2 are not observed in non-hydrogenated N-doped GaAs samples. B1/B2 and C1/C2 show similar depressions of the surface as N atoms observed before, but they are decorated with a bright contrast in different crystallographic directions. These features will be described in more detail below. There are more types of features present in this image, but these show up less regularly than A, B1/B2 and C1/C2. Because of this, we will not discuss these features in this manuscript. The large amount of different features present make it impossible to verify the nominal concentration of 0.62\%. But based on this concentration, 26$pm$1\% of the N atoms appear in configuration B1/B2 and 33$pm$1\% of the N atoms appear in configuration C1/C2. Due to the inherent instability of the features described in this manuscript it was not possible to perform high quality scanning tunneling spectroscopy (STS) measurements on them. Performing a single STS measurement on a feature easily affects the feature, as a result it was not feasible to obtain reliable STS results on these features.
    
    \subsection{Comparison between the annealed and non-annealed sample}
    
        \begin{figure}
            \centering
            \includegraphics{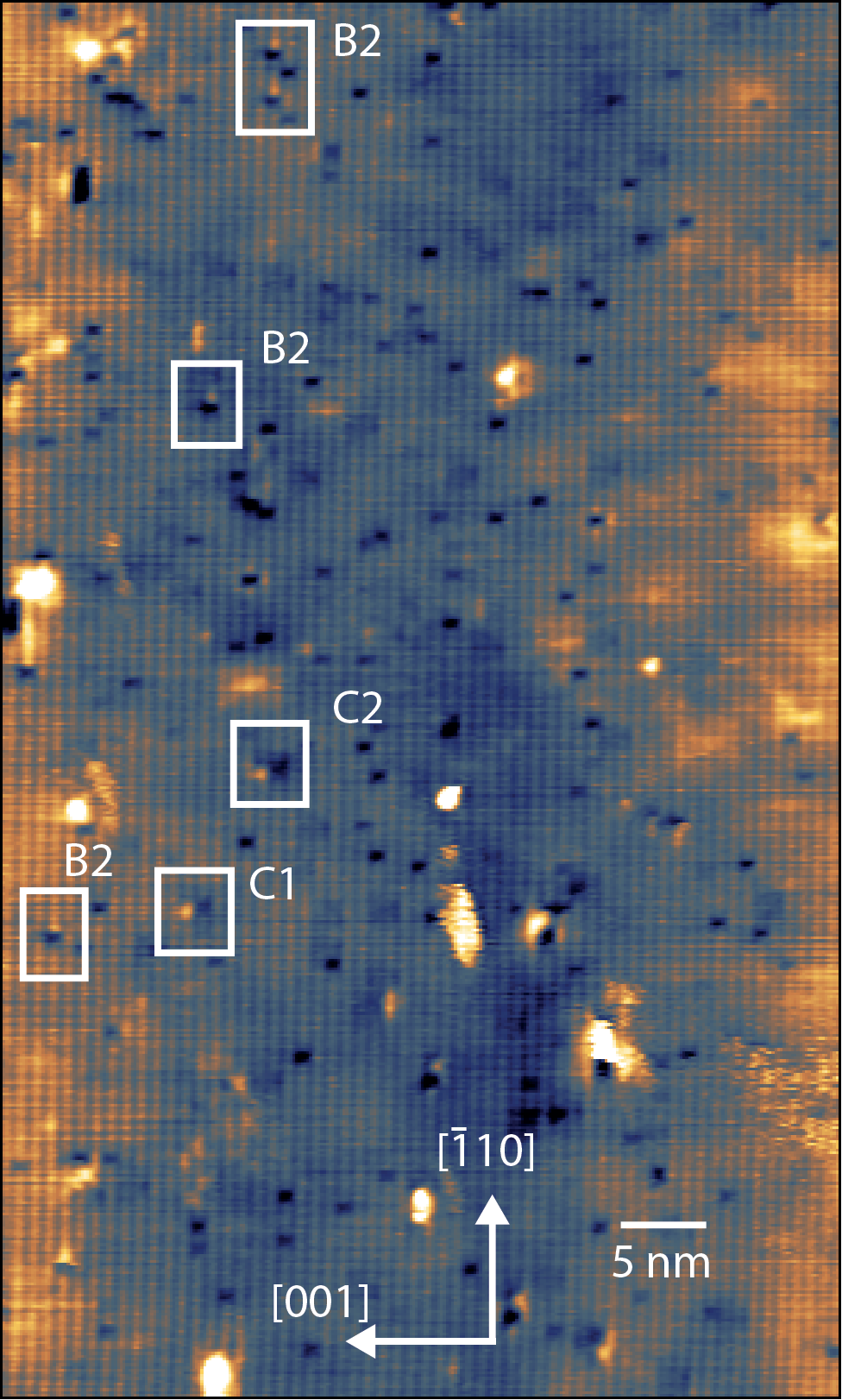}
            \caption{Filled state image of the annealed sample obtained with a sample bias of -3 V and a tunnel current of 30 pA. Decorated N atoms are marked with white rectangles and an indication of their type.}
            \label{fig:annealedstm}
        \end{figure}
       
        Measurements performed on the annealed sample, which should contain primarily N-2H complexes, show almost none of these new features A, B1/B2 and C1/C2. This can be seen in Fig. \ref{fig:annealedstm}, where the decorated N atoms present in the image are marked with white rectangles. Of all the N atoms present in this image, only 6$\pm$1\% appear in configuration B1/B2 or C1/C2. The rest of the N atoms in the image exhibits the characteristics of non-hydrogenated N atoms.
        
       One could suggest that N-2H complexes do not show any contrast in the STM measurements. Because the N-2H complex is strain free, it could mean that the (110) surface is undisturbed at the location of the N-2H complex. This possibility is however ruled out by checking the observed concentration of N features in an image of the annealed sample. The sample has a nominal N concentration of 1\%, as estimated by XRD. The concentration of N atoms observed is 0.94$\pm$0.02\% and an additional 0.06$\pm$0.01\% of the atoms are decorated N atoms in the B1/B2 or C1/C2 configuration. So the observed value of N atoms is almost the same as the expected value from the growth menu. Which could mean that N-2H shows exactly the same characteristics as a non-hydrogenated N atom, but we will argue why this is unlikely. The N features in the annealed sample show all the characteristics of undecorated N atoms, such as the depression of the surface in filled state imaging and the bowtie-like shape in empty state imaging. Because the N-nH complexes change the bandgap of N-doped GaAs it is likely that they also affect the electronic state of the N atom, but we do not observe that effect here. Therefore we propose that the N-2H complex is not stable near the (110) surface of GaAs, meaning that the 2 H atoms of the N-2H complex dissociate and disappear as H$_2$ into the vacuum. This would leave the N atoms near the (110) surface, which are the ones we can observe with the STM, without any H atoms, resulting in the observation of undecorated N atoms. In the next section we discuss this instability in more detail, including the process where N-3H complexes near the surface reduce to N-H. This means that in the non-annealed sample we would observe only N-H complexes, whereas in the annealed sample only non-hydrogenated N atoms would be visible. 
        
    \subsection{DFT calculations of N-nH complexes}
    \label{sec:DFT}
        \begin{figure}
            \centering
            \includegraphics{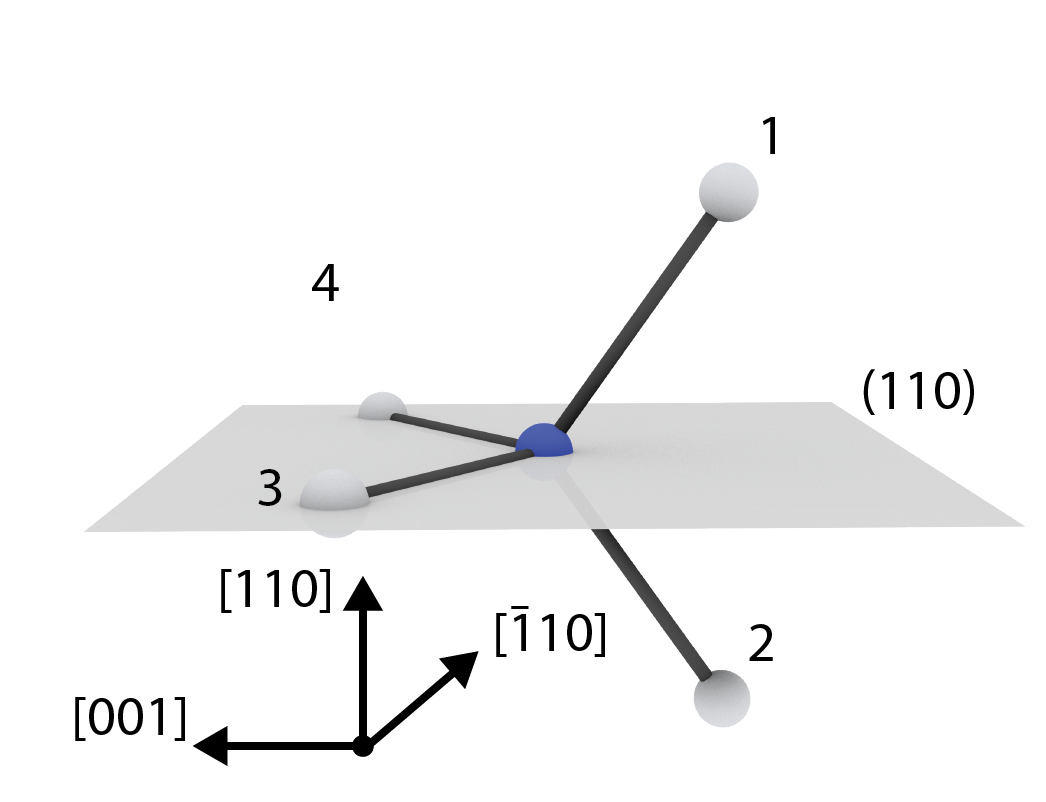}
            \caption{N atom (blue) in the (110) surface of a III/V semiconductor. The four bonds the N atom has to neighboring atoms are marked with the number 1 to 4. Bond 1 is a dangling bond directed out of the (110) surface. Bond 2 connects the N atom to an atom in the layer below the surface layer. Bond 3 and 4 connect the N atom to neighboring atoms in (110) surface.}
            \label{fig:N4H}
        \end{figure}
        
        Based on the suggestions of the previous section, we simulated the behavior of N-nH complexes near the GaAs (110) surface using DFT. All these simulations involve a single N atom in the (110) surface of GaAs, as schematically depicted in Fig. \ref{fig:N4H}. The N atom (blue) has bonds to four neighboring sites, marked with the numbers 1 to 4. Bond 1 is a dangling bond directed out of the (110) surface. Bond 2 connects the N atom to an atom in the layer below the surface layer. Bond 3 and 4 connect the N atom to neighboring atoms in (110) surface. Hereafter we will use this code to refer to the different N bonds. The lattice orientation in the DFT results following below is the same as in Figure \ref{fig:N4H}.
        
        \begin{figure}
            \centering
            \includegraphics{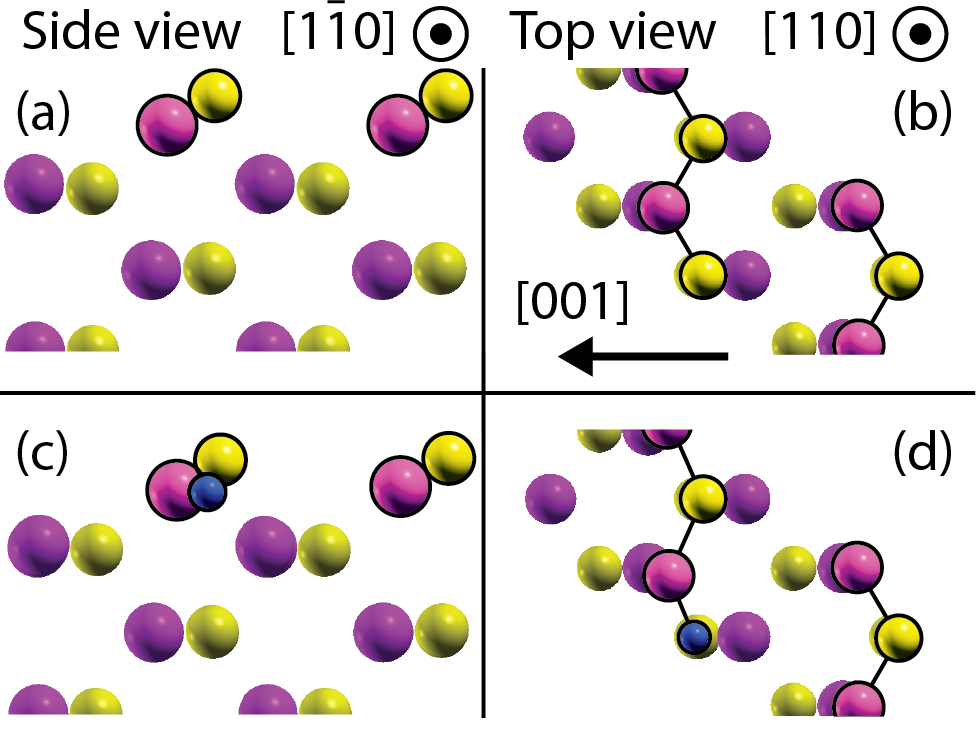}
            \caption{DFT calculations of a relaxed clean GaAs slab as seen from a side view (a) and a top view (b) of the (110) surface. Ga and As atoms are colored purple and yellow respectively. The bottom row shows the relaxed GaAs slab with a substitutional N atom (colored blue) in the first layer of the (110) surface as seen from a side view (c) and top view (d). The solid circles and lines indicate the atoms in the surface corrugation rows.}
            \label{fig:dft_gaas}
        \end{figure}
        
        To test the DFT model we first calculated the relaxations of a clean GaAs slab, as shown in Fig. \ref{fig:dft_gaas}. The top row of Fig. \ref{fig:dft_gaas} shows the (110) surface from a side view (a) and a top view (b). Here the characteristic buckling of the (110) surface can be observed, where the As atoms move outwards and the Ga atoms inwards which is well known for the (110) surface of GaAs \cite{lubinsky1976semiconductor,chadi1979110,kahn1986atomic}. The bottom row of Fig. \ref{fig:dft_gaas} shows the (110) GaAs surface with a substitutional N atom on an As site on the surface. After relaxation the N atom moves into the surface and the surrounding Ga atoms are slightly displaced towards it, as can be seen in Fig. \ref{fig:dft_gaas}(c) and (d) from a side and top view respectively. This relaxation is typical for the N atom in the surface, as was shown in Ref \cite{Tilley2016ScanningModel}. 
        
        \begin{figure*}
            \centering
            \includegraphics{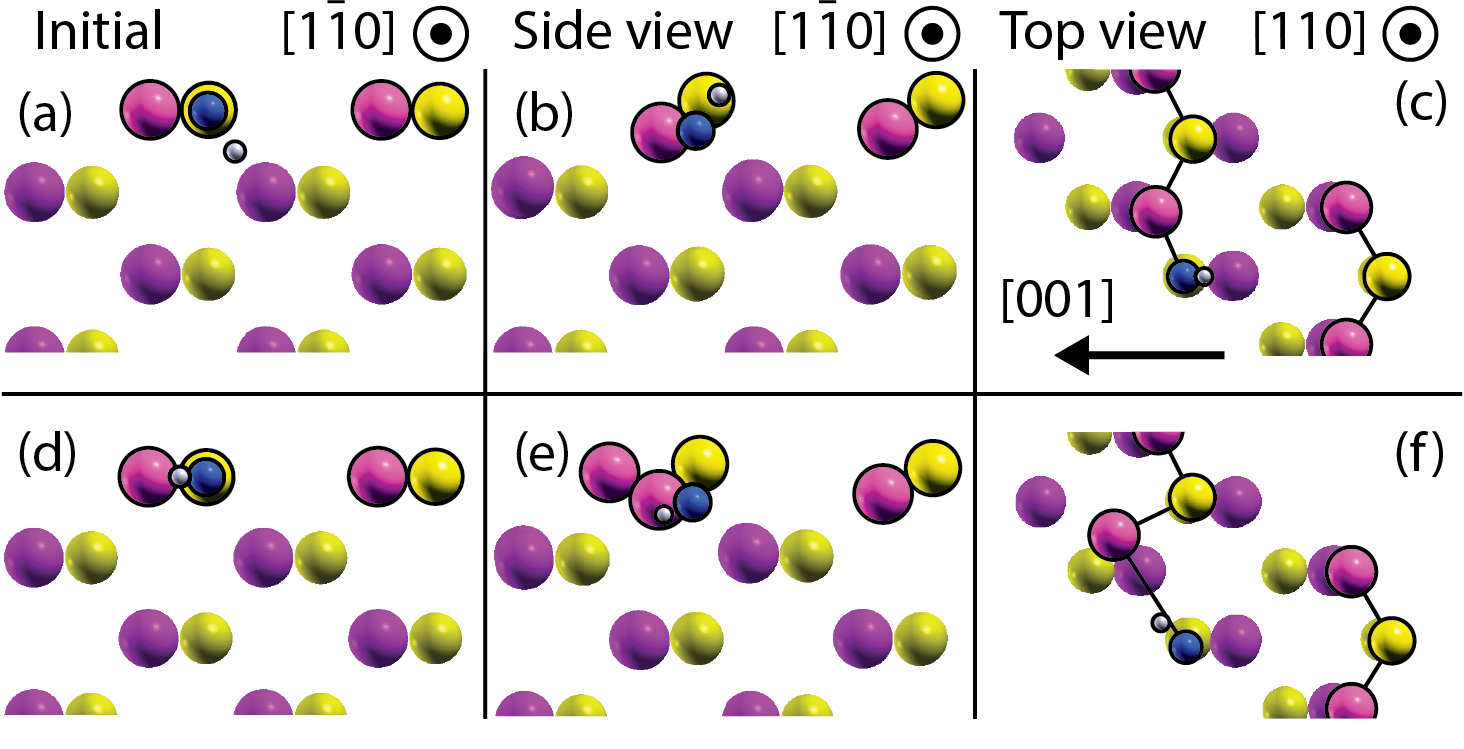}
            \caption{DFT calculations of a single H atom (white) bound to an N atom (blue) in the (110) surface of GaAs. Ga and As atoms are colored purple and yellow respectively. Two different initial states are displayed in (a) with the H atom located along bond 2 and (d) with the H atom oriented along bond 3 or 4. (b)(c) show the orientation of (a) in a relaxed situation from a side (b) and top (c). (e)(f) show the orientation of (d) in a relaxed situation from a side (e) and top (f) view. The solid circles and lines indicate the atoms in the surface corrugation rows.}
            \label{fig:dft_1H}
        \end{figure*}
        
         We now extend the model with H atoms in close vicinity to the N atom to model N-nH complexes. The results of these calculations for N-1H complexes are displayed in Fig. \ref{fig:dft_1H}. Fig. \ref{fig:dft_1H}(a) displays an initial state where the H atom is oriented into the surface halfway along bond 2 between the N atom and the Ga atom. After the relaxation it can be observed in Fig. \ref{fig:dft_1H}(b) (side view) and (c) (top view) that the H atom has oriented itself out of the surface along bond 1. Fig. \ref{fig:dft_1H}(c) displays a second initial state of the system where the H atom is oriented along bond 3 or 4 (which are mirror-symmetric with respect to the [001] direction). After the relaxation the H atom is still primarily oriented in this direction, as can be seen in Fig. \ref{fig:dft_1H}(e) (side view) and (f) (top view).
        
        \begin{figure}
            \centering
            \includegraphics{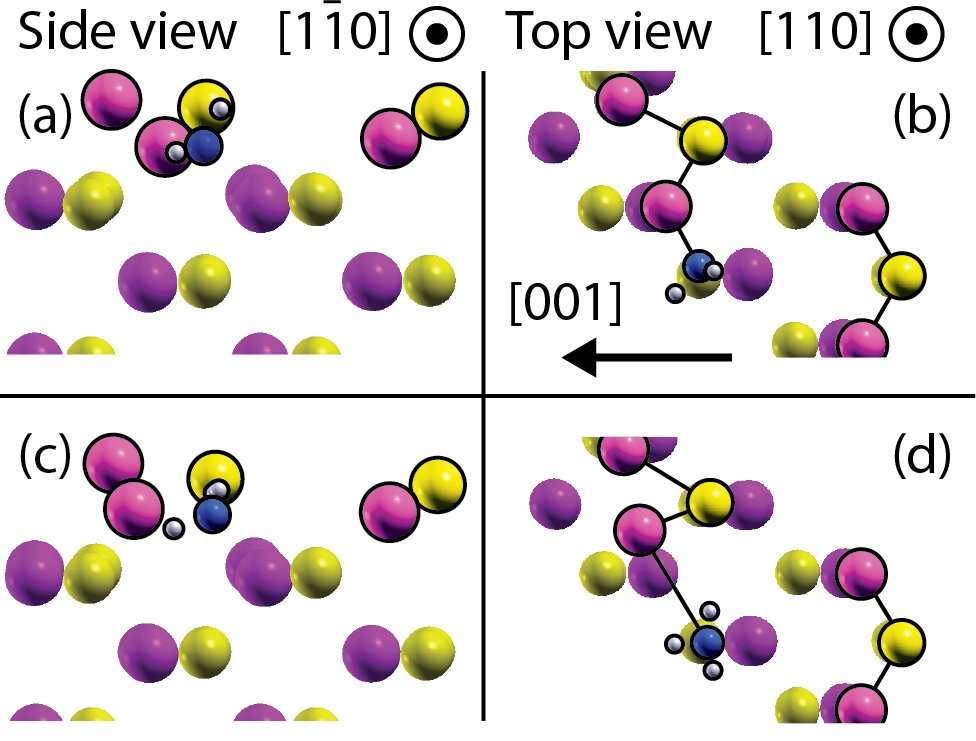}
            \caption{DFT calculations of N-nH complexes in a GaAs lattice. The N atom (blue) is located in the top layer of the (110) plane of GaAs with H atoms (white) in the vicinity. Ga and As atoms are colored purple and yellow respectively. The top row shows a relaxed N-2H complex from a side view (a) and a top view (b) orientation. The bottom row shows a relaxed N-3H complexes from a side view (c) and a top view (d) orientation. The solid circles and lines indicate the atoms in the surface corrugation rows.}
            \label{fig:dft_23H}
        \end{figure}
        
        For N-2H complexes we observe that both of the H atoms orient themselves out of (bond 1) or parallel (bond 3 or 4) to the surface even if they were originally oriented into the crystal (band 2). This is schematically displayed in Fig. \ref{fig:dft_23H}(a) and (b) where one H atom is located at bond 1 and one is located at bond 3, as seen from a side view (a) and a top view (b). Here we observe that the Ga atom that is located in the direction of the H atom shows a large displacement away from the N atom and out of the surface. This displacement should be visible in STM, but as shown before in the annealed sample (which should contain mainly N-2H) we observe only N atoms that in all aspects look like normal non-hydrogenated N atoms. This provides credibility to our proposal that the two H atoms dissociate from N-nH complexes as H$_2$. Another aspect to consider is the energy associated with the different configurations. The activation energy of the N-2H complex has been estimated at 1.89 eV \cite{Bisognin2008High-resolutionNitrides}. Since a hydrogen molecule has a binding energy of 4.52 eV \cite{cottrell1958strengths}, the dissociation of the two H atoms from the N-2H complex is energetically favorable. We propose that this process is unlikely in the bulk of the crystal since there is no space for the H$_2$ molecule to go, but at the surface the H$_2$ molecule can easily move into the nearby vacuum. 
        
        A similar process could occur for N-3H complexes. For these complexes DFT calculations show that two H atoms relax outwards towards bond 1 while the third one remains in the (110) plane as can be seen in Fig. \ref{fig:dft_23H}(c) from a side view and (d) from a top view. Again based on the activation energy, which is estimated to be 1.77 eV for N-3H \cite{Bisognin2008High-resolutionNitrides}, we propose that 2 of the H atoms dissociate as an H$_2$ molecule. This would mean that the observed features related to hydrogenated N originate from N-1H complexes. 

    \subsection{A, B1/B2 and C1/C2 features}
        Here we will give an extensive description of the features A, B1/B2 and C1/C2 taking into account the theoretical predictions of the previous section. The lattice orientation of the features depends on the specific (110), or equivalent, plane revealed after cleaving. If the lattice orientation of the wafer is not known before sample preparation it is unknown which (110) plane will be revealed after cleaving. 
        The characteristic bow-tie like shape of the N atom at empty state imaging conditions shows an clear anisotropy in its contrast. Using this anisotropy and comparing our measurements with Refs \cite{Plantenga2017SpatiallyGaAs,Ishida2015DirectMicroscopy} we were able to determine the exact lattice orientation of the surface. \\
        
        \subsubsection{A Feature}
        Feature A has a bright two-lobed structure as seen in Figure \ref{fig:features}. The two lobes of the structure are located at neighboring As lattice sites. The long axis is oriented along to the [110] crystal axis, while the short axis is oriented perpendicular to this, in the [001] direction. In the [00$\Bar{1}$] direction between the two lobes, a small reduction of the LDOS is observed at a location between the corrugation lines. 

         These features are mobile when scanning at slightly larger negative voltages (-3.4 V instead of -3 V). In empty state imaging tunneling is suppressed near these features. We classify these features as Ga vacancies based on similar observations made by Domke \textit{et al} \cite{Domke1996MicroscopicGaAs} and Lengel \textit{et al} \cite{Lengel1996ChargeGaAs110}. The empty dangling bond of the Ga atom is absent and the occupied dangling bonds of the As atoms neighboring it are raised. Similar observations on N-doped GaAs were reported by Ishida \textit{et al} \cite{Ishida2015DirectMicroscopy}. They observed these features only sporadically, whereas we observe 194 of these features in a 135$\times$50 nm$^2$ area. We attribute this higher occurrence of Ga vacancies to the effects of the hydrogenation. Since the hydrogenation breaks N-Ga bonds, the atoms in the outermost layer of the surface can then be (depending on the Ga bonds that have been broken by the hydrogenation) less tightly bound to the bulk crystal. These weaker bonds could in turn lead to more Ga vacancies. The DFT calculations of the N-nH complexes show a large outward displacement of one of the Ga atoms next to the complex. This could cause Ga atoms to leave their original position in the lattice, resulting in an increase in Ga vacancies in the hydrogenated N-doped layer. As a result of the mobility of these vacancies they do not need to reside next to the N atom where they initially formed during the cleaving. \\

        \subsubsection{B1/B2 Feature}
        Feature B1/B2 consists of an apparent decrease of the local density of states (LDOS) at a single corrugation site (most likely due to a physical depression), similar to a single N atom in the surface layer, with an additional localized apparent enhancement of the LDOS neighboring it. This can be observed in Figure \ref{fig:features}.
        Both the decrease and enhancement of the LDOS are located on the same corrugation line. The bright contrast is located in either the [1$\Bar{1}$0] or the [$\Bar{1}$10] crystal direction for feature B1 or B2 respectively. In a 135x50 nm area of the non-annealed sample we observed 40 occurrences of B1 and 41 of B2. The fact that B1 and B2 are observed with equal frequency strongly points to the possibility that B1 and B2 are equivalent isomeric features. 
        
        \begin{figure}
            \centering
            \includegraphics{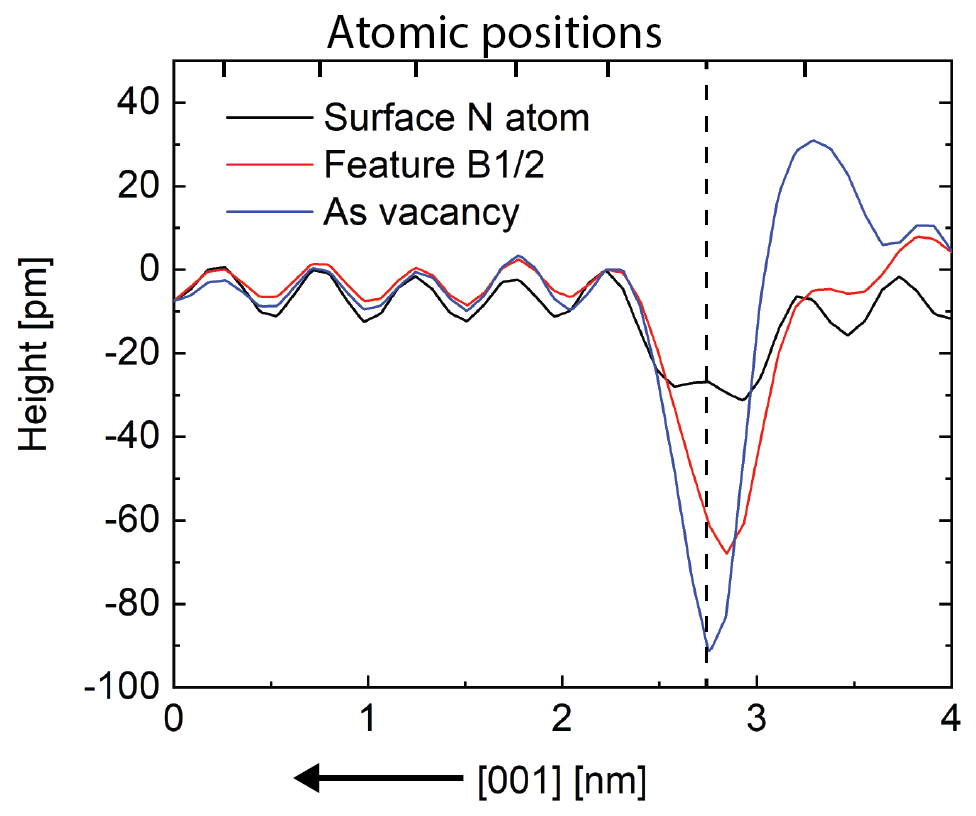}
            \caption{Height profiles along the [001] direction of a surface layer N atom, feature B1/B2 and an As vacancy. All profiles were obtained in the non-annealed sample by taking the average profile of three features. The atomic corrugation to the left of the features is used to align the profiles. The dotted line marks where the original corrugation maximum would be if the features would not be present. Heigth profiles were taken from images obtained with a sample bias of -3.35 V and a tunnel current of 30 pA. }
            \label{fig:height}
        \end{figure}
        
        Fig. \ref{fig:height} displays the height profiles of feature B1/B2, an undecorated surface layer N atom and an As vacancy. The dotted line marks where the original corrugation maximum would be if the features would not be present. The atomic corrugation to the left of the features is used to align the profiles. All the profiles are obtained by taking the average height profile of three features, to prevent local perturbations from affecting the profiles in the figure. The surface layer N atom exhibits a depression of 30 pm compared to the top of the atomic corrugation, which is similar to the magnutide reported by Ulloa \textit{et al} \cite{Ulloa2008StructuralMicroscopy}. Feature B1/B2 exhibits a depression of 70 pm and the As vacancy a depth of 90 pm. The depth of the depression exhibited by feature B is in between that of N in the surface layer and the atomic vacancy. We suggest that feature B1/B2 is an N atom which has been pulled deeper into the surface as a result of the hydrogenation. 
        
        Noting the mirror symmetric nature with of feature B1 and B2 respect to the [001] axis, we suggest that the feature involves an N atom with an H atom on either position 3 or 4 (as defined in Section \ref{sec:DFT}) since these positions have the same mirror symmetry. A single H atom on position 3 or 4 causes the N atom to relax into the surface by an additional 1 to 5 pm in the DFT calculations. The magnitude of the inward relaxation of the surface N atom due to the hydrogen atom is much smaller in the DFT calculations. Note that the normal surface N atom in the DFT calculations already relaxes 82 pm into the surface, which is similar to the value found by Tilley \textit{et al} \cite{Tilley2016ScanningModel}, but much larger than the experimentally found value.
        
        \begin{figure}
            \centering
            \includegraphics{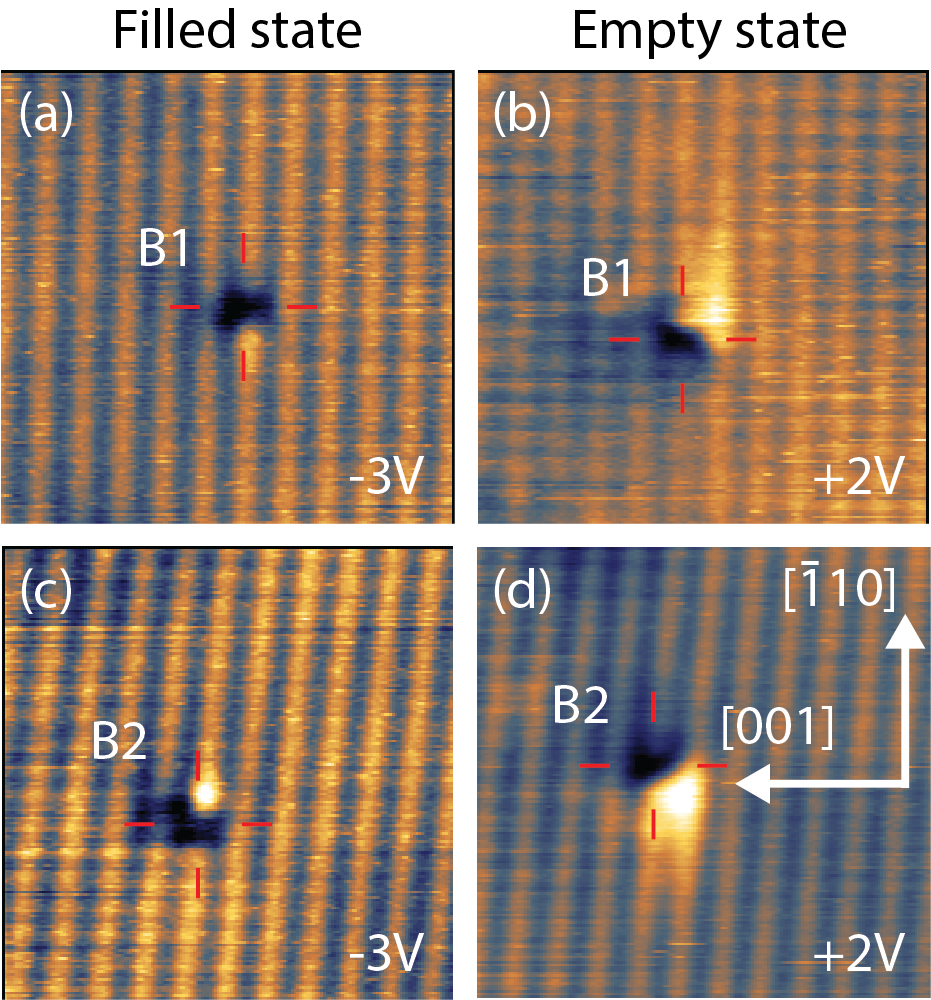}
            \caption{5$\times$5 nm$^2$ STM images of feature B1 (a) (b) and B2 (c) (d). Filled state images a) and b) were obtained at a bias voltage of -3 V, while c) and d) are empty state images were obtained at +2 V and a tunnel current of 50 pA. The location of the N atom in the features is marked with a red cross.}
            \label{fig:Bposneg}
        \end{figure}
        
        Fig. \ref{fig:Bposneg} displays a set of filled (a,c) and empty (b,d) state images of feature B1 (a,b) and B2 (c,d). At empty state tunneling voltages the depression of the surface, as observed in filled state measurements, is still present. In addition to this we observe an apparent enhancement of LDOS spread over two corrugation rows. The brightest part is located next to the depression and it decays over a distance of $\sim$2 nm in the direction opposite to the bright dot observed in the filled state image. At lower positive voltages the bright part of the feature disappears, indicating that it is related to an electronic state contributing to the contrast at a specific energy. 
        The contrast caused by an N atom in the surface layer of the (110) plane has been observed and described as an enhancement of the LDOS in the [110] direction spread over two corrugation rows, with the N atom located in the middle \cite{Plantenga2017SpatiallyGaAs,Ishida2015DirectMicroscopy}. The feature we observe shows strong similarity to this feature, with the major difference that the contrast only spreads from the N atom in one crystal direction instead of two. We explain this behavior by considering the effect of the hydrogenation. If the H atom is located on bond 3 or 4 that originally connected the N atom to a Ga atom in the same corrugation row, the N atom will have a reduced coupling with that Ga atom, and as a result the wavefunction and its related state would not spread in that direction. In the DFT calculations we also observe that the Ga atom that is located in the direction of the H atom moves away from the N atom, as shown in Figure \ref{fig:dft_1H}(d-f), further reducing the coupling. We attribute the observed enhancement of the LDOS in feature B to the raising of the filled dangling bond of the As atom located in the same direction as the H atom, so in the direction of bond 3 or 4. The main indication for this is the fact that this enhancement is not observed in the empty state images of feature B. \\
        
        \subsubsection{C1/C2 Feature}
        
        \begin{figure}
            \centering
            \includegraphics{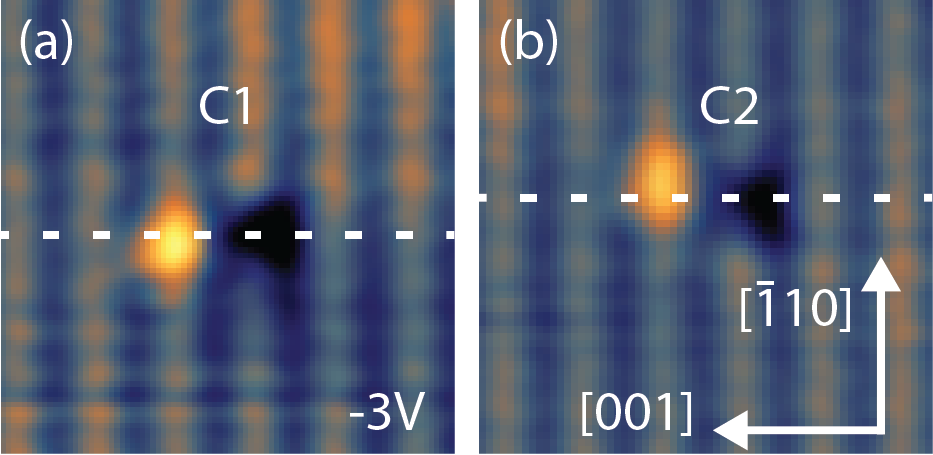}
            \caption{3$\times$3 nm$^2$ images of feature C1 (a) and C2 (b). Dotted lines are drawn through the center of the N atoms parallel to the [001] direction to illustrate the displacement of the apparent enhancement of the LDOS. Both images were taken with a bias voltage of -3.35 V and the tunnel current set to 30 pA.}
            \label{fig:C1C2}
        \end{figure}
        
        In filled state images such as Figure \ref{fig:features}, feature C1/C2 shows an apparent reduction of the LDOS in the surface corrugation with an apparent enhancement of the LDOS in the neighboring corrugation line in the [00$\bar{1}$] direction. This enhancement has a slight displacement in the [1$\bar{1}$0] (C1) or [$\bar{1}$10] (C2) direction relative to the depression of the surface as can be seen in Figure \ref{fig:C1C2}(a) and (b) respectively. C1 and C2 are observed 48 and 54 times respectively in a 135$\times$45 nm$^2$ area. The apparent reduction of the LDOS has almost the same characteristic depth as an undecorated N atom in the first layer of the surface. 
        
        \begin{figure}
            \centering
            \includegraphics{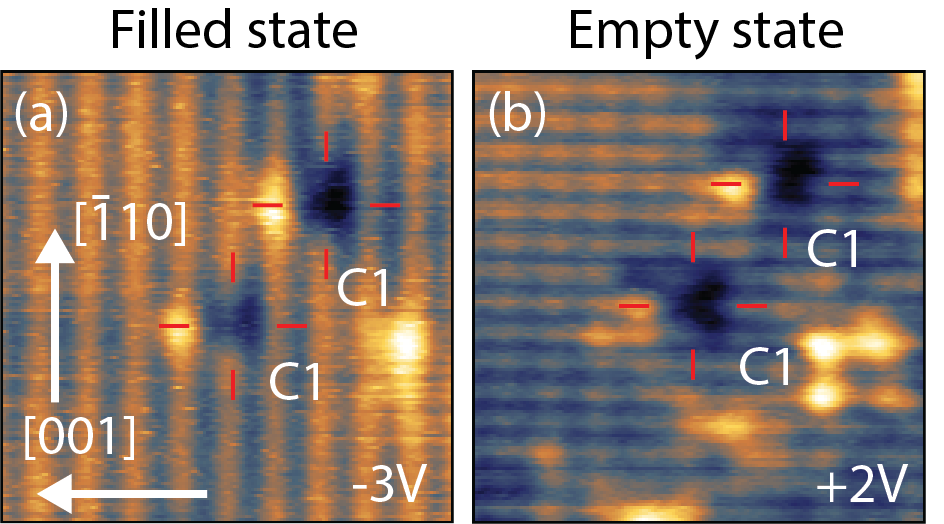}
            \caption{5$\times$5 nm$^2$ images of two instances of feature C1 at a bias voltage of a) -3 V and b) +2 V. Both images were taken with the tunnel current set to 30 pA. }
            \label{fig:Cposneg}
        \end{figure}
        
        Fig. \ref{fig:Cposneg} shows a filled a) and empty b) state image of two instances of feature C1. Empty state images of feature C still show the depression of the surface. The bright dot is also still visible. The main difference observed is a light depression around the main dark spot at the location of the N atom. This light depression extends about 2 corrugation rows in all directions. The difference between the filled and empty state image is quite small, indicating that the feature is probably almost completely topographic in origin. 

        Looking at the statistics, in the same 135$\times$45 nm$^2$ area we observe feature B1 40 times, B2 41 times, C1 48 times and C2 54 times. If we assume that all the bonds of the N atom have an equal chance to be terminated by an H atom during the hydrogenation process, we should observe similar numbers of feature B1, B2, C1 and C2, which is the case. We suggested that feature B has the H atom bonded along bond 3 or 4, as defined in Section \ref{sec:DFT}. Following our suggestion that these features involve a single N and H atom, this would mean that feature C1/C2 involves bond 1 or 2.  
            
        Bond 1 and 2 are both oriented in the [00$\bar{1}$] direction, with bond 1 pointing out of the (110) surface and bond 2 facing into the bulk. In the DFT calculations we observe that an H atom initially placed on bond 2 relaxes out of the surface and moves to a location on bond 1. An H atom initially placed on bond 1 stays in that position. Based on this we suggest that at the (110) surface N atoms with an H atom at either bond 1 or bond 2 will relax towards the same position. In the experiment the surrounding crystal can cause symmetry breaking, resulting in the formation of the isomeric features C1 and C2. We attribute this apparent enhancement of the LDOS to physical upwards movement of an As and Ga atom in the neighboring corrugation row, since the apparent enhancement is observed in both filled and empty states images. 
    
    \subsection{Manipulation and stability of feature B1/B2 and C1/C2}
        N-nH complexes are known to dissociate at elevated temperatures \cite{Berti2007FormationNitrides} or when exposed to laser light of high intensity \cite{Balakrishnan2012Band-gapIII-N-Vs}. Both of these methods rely on the injection of energy into the complexes to break them. With the STM it is also possible to supply energy to the complex, this can be done by performing voltage ramps with the feedback loop disabled. In this way currents of up to 3 nA can be injected into the sample at the location of the STM tip. In addition, the STM tip can be brought closer to the surface to change the localization of the current. 
        
         \begin{figure}
            \centering
            \includegraphics{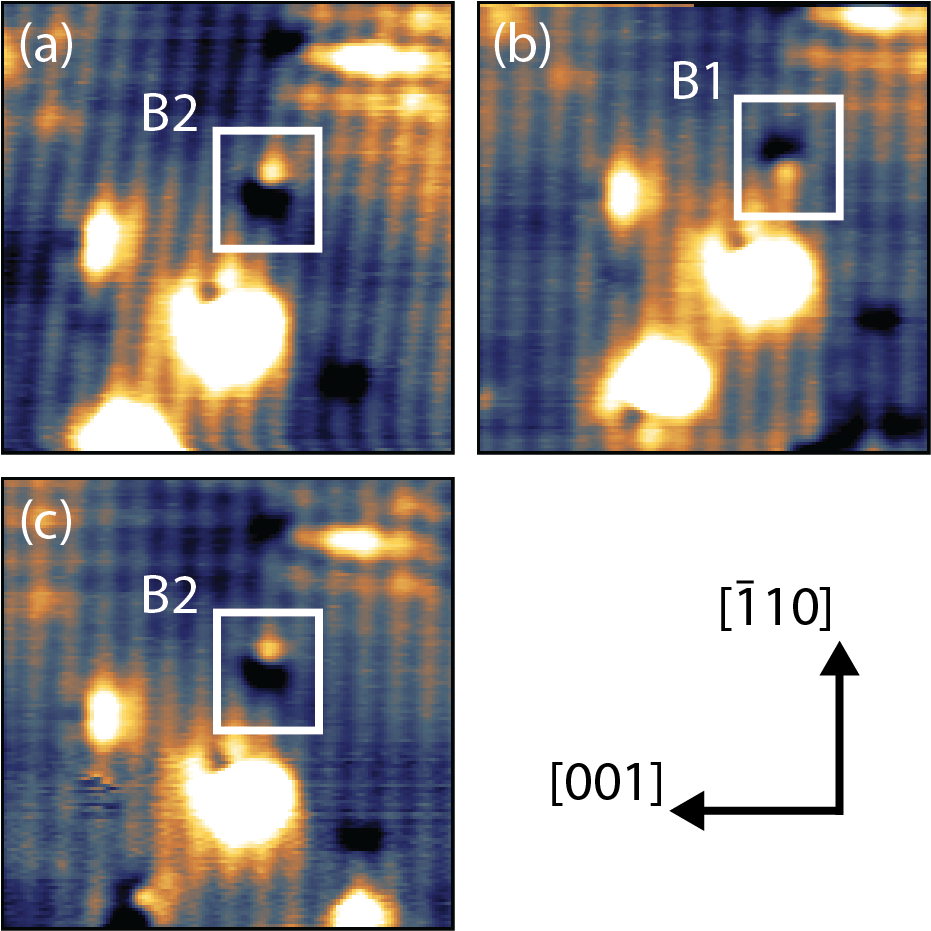}
            \caption{6.5$\times$6.5 nm$^2$ filled state images (-3 V, 30 pA) of the annealed sample containing an instance of feature B1/B2. Between each image a voltage ramp was performed at the location of feature B1 or B2, which is marked with a white rectangle. The image series shows switching of the feature from B2 a) to B1 b) and back to B2 again c) while the surrounding area of the sample remains unchanged.}
            \label{fig:Bswitching}
        \end{figure}
        
        Performing voltage ramps at the location of feature B1/B2 very often has no effect, the contrast profile remains at the same position in the lattice and retains its shape. In one case we observed that feature B1/B2 could be made to switch back and forth between orientation B1 and B2 when performing voltage ramps at the location of the feature as illustrated in Fig. \ref{fig:Bswitching}.  It was not possible to remove the bright contrast. This switching behavior supports the isomeric nature of the B1 and B2 complexes involving bond 3 and 4 as defined in Section \ref{sec:DFT}.
        
         \begin{figure}
            \centering
            \includegraphics{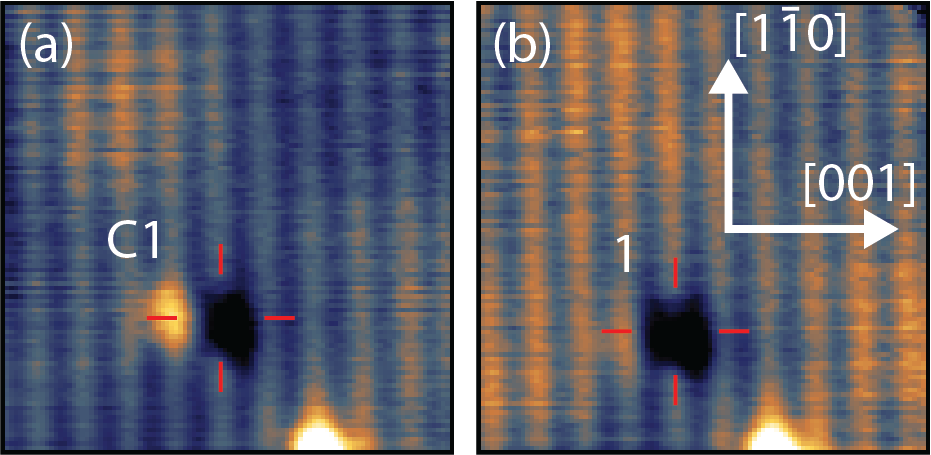}
            \caption{The LDOS enhancement part of feature C can be removed by performing voltage ramps at the location  of this feature. Image a) and b) are subsequent images (4.5$\times$4.5 nm$^2$) of the same location, in between the images a voltage ramp from -3 V to +0.5 V was performed with the feedback loop off.} 
            \label{fig:c_removal}
        \end{figure}
        
        Voltage ramps were also performed at the position of feature C1/C2. The results of this are displayed in Fig. \ref{fig:c_removal}. After taking the voltage ramp and scanning the same area again, the apparent localized enhancement of the LDOS has disappeared, leaving behind a normal undecorated first layer N atom. The depression of the surface is on average 5 pm deeper after removal of the apparent enhancement of the LDOS. This could either be caused by a physical change in the depth of the feature, or because the tip can follow the depression better after the removal, causing the feature to appear deeper. This deepening of the feature is also observed in the DFT calculations, where the N atom with an H atom located in the direction of bond 2 has relaxed around 10 pm towards the surface compared to a non-hydrogenated N atom. Performing more voltage ramps at the location of the depression once the LDOS enhancement has disappeared does not make it re-appear. Switching between configuration C1 and C2 was not observed, since attempt at manipulation always resulted in the removal of the apparent enhancement of the LDOS. Since the removal of the LDOS enhancement is irreversible and the N atom involved has the characteristics of a non-hydrogenated N atom after the removal, it is very likely that the disappearance of the LDOS enhancement of feature C1/C2 is caused by the removal of an H atom bound to the N atom. This would mean that it is possible to manipulate the hydrogenation of N-doped GaAs at the atomic scale. 

\section{Conclusions}
    We performed X-STM measurements on single N atoms in hydrogen irradiated N-doped GaAs samples. On these samples we observed features related to the hydrogenation process, of which a subset is classified as N-nH complexes. There are strong indications that two of the H atoms of N-nH (n$\geq$2) complexes dissociate from the crystal as H$_2$ when the complexes reside at or near the (110) surface of GaAs. This is further supported by DFT calculations showing a preferred outwards orientation of the H atoms. The features are studied in both filled and empty state imaging, providing information on their topographic and electronic nature. 
    
    The observed N-H features can also be manipulated by performing voltage ramps with the STM tip at the location of these features. For feature B1/B2 this was shown as a switching behavior between the two mirrored orientations, implying an isomeric relationship between the two crystal orientations of B1 and B2. For feature C1/C2 manipulation consisted of a removal of the bright contrast, resulting in a feature identical to a non-hydrogenated N atom. This indicates that the process involved in this manipulation is likely the removal of the H atom from the N-H complex. In conclusion, this work clarifies some aspects of the physics of N-nH complexes in hydrogenated N-doped GaAs, paving the way towards a higher control of the properties of these materials for photonics and quantum information technology applications.\\

\begin{acknowledgments}
    We acknowledge S. Rubini (IOM-CNR, Italy) for providing samples at an early stage of the research. A.G. and J.M.U. acknowledge funding from the Spanish MINECO through project MAT2016-77491-C2-1-R. M.S., M.F., and A.P. acknowledge funding from the European Union’s Horizon 2020 Research and Innovation Program under the Marie Skłodowska-Curie Grant Agreement No. 641899 (PROMIS), from the Regione Lazio Program “Progetti di Gruppi di ricerca” legge Regionale n. 13/2008 (SINFONIA project, prot. n. 85-2017-15200) via LazioInnova spa, and from Sapienza Università di Roma via Fondi Ateneo 2018-2019, SapiExcellence, and Avvio alla Ricerca. C. S. and M. E. F. acknowledge support from the NSF DMREF program through Award No. DMR-1921877.
\end{acknowledgments}

\providecommand{\noopsort}[1]{}\providecommand{\singleletter}[1]{#1}%

\end{document}